\documentclass[aps,pra,groupedaddress,notitlepage]{revtex4-1}

\usepackage{amssymb}
\usepackage{amsmath}
\usepackage{amsfonts}
\usepackage{amsthm}
\usepackage{graphicx}
\usepackage{color}

\usepackage{url}
\newcommand{\R}{{\mathbb R}}
\newcommand{\E}{\mathbb{E}}

\begin{document}

\title{Multiscale and Anisotropic Characterization of Images Based on Complexity: an Application to Turbulence}

\author{Carlos~Granero-Belinchon$^{1,2}$}
\email{carlos.granero-belinchon@imt-atlantique.fr}
\author{St\'ephane G. Roux$^{3}$}
\email{stephane.roux@ens-lyon.fr}
\author{Nicolas B. Garnier$^{3}$}
\email{nicolas.garnier@ens-lyon.fr}
\affiliation{
$^{1}$ Department of Mathematical and Electrical Engineering, IMT Atlantique, Lab-STICC, UMR CNRS 6285, 655 Av. du Technop\^ole, Plouzan\'e, 29280, Bretagne, France. \\
$^{2}$ Odyssey, Inria/IMT Atlantique, 263 Av. G\'en\'eral Leclerc, Rennes, 35042, Bretagne, France. \\
$^{3}$ Univ Lyon, ENS de Lyon, CNRS, Laboratoire de Physique, 46, all\unexpanded{\'e}e d'Italie, Lyon, F-69364, Rh\unexpanded{\^o}ne-Alpes, France}           
            
\begin{abstract}
This article presents a multiscale, non-linear and directional statistical characterization of images based on the estimation of the skewness, flatness, entropy and distance from Gaussianity of the spatial increments. These increments are characterized by their magnitude and direction; they allow us to characterize the multiscale properties directionally and to explore anisotropy. To describe the evolution of the probability density function of the increments with their magnitude and direction, we use the skewness to probe the symmetry, the entropy to measure the complexity, and both the flatness and distance from Gaussianity to describe the shape. These four quantities allow us to explore the anisotropy of the linear correlations and non-linear dependencies of the field across scales. First, we validate the methodology on two-dimensional synthetic scale-invariant fields with different multiscale properties and anisotropic characteristics. Then, we apply it on two synthetic turbulent velocity fields: a perfectly isotropic and homogeneous one, and a channel flow where boundaries induce inhomogeneity and anisotropy. Our characterization unambiguously detects the anisotropy in the second case, where our quantities report scaling properties that depend on the direction of analysis. Furthermore, we show in both cases that turbulent velocity fluctuations are always isotropic, when the mean velocity profile is adequately removed.
\end{abstract}

\maketitle 

\section{Introduction}

Nowadays image processing is a fundamental step in the quantitative study of a large number of domains such as: remote sensing~\cite{Du2002,IsernFontanet2007}, medicine~\cite{Castellano2004, Gerasimova2014}, physics~\cite{Everson1990,Yeung2015} or ecology~\cite{Stepanian2014}. Very commonly, experimentally acquired signals or numerical simulation results are viewed as composed of a mean behavior over which are superimposed fluctuations considered as a stationary or homogeneous stochastic field. For complex systems, this stochastic field generally results from non-linear dynamics and exhibits multiscale behavior~\cite{Lumley1970,Mohtar2018,Brandenberger1992,Robert2008}. Moreover, for multidimensional signals, for example images, volumes or matrices of any dimensionality, the mean behavior as well as the fluctuations can be anisotropic, \textit{i.e.} their behavior may depend on the direction of analysis. So, the characterization of images or multi-dimensional signals needs methodologies able to deal with multiscale, non-linear and anisotropic properties. In this work we study images and we focus on their fluctuations considered as a 2D stochastic field.

Current multiscale statistical analyses of 2D fields are mainly based on second order statistics such as 2D Fourier transform and 2D auto-correlation~\cite{Du2002,Everson1990,Garnier2002,Wang2020,Deshpande2000}. They are able to provide direction-dependent descriptions of linear dependencies across scales. For example, a direction-dependent roughness characterization of Gaussian fields has been proposed in~\cite{Richard2016,Richard2018, Vu2020}. However for non-Gaussian fields, all methods cited above are blind to non-linear dependencies. Consequently, other image processing techniques such as multifractal analysis, high-order statistical moments or cumulants were also developed in the last decades~\cite{IsernFontanet2007,Everson1990,Maussang2007,Dimitrakopoulos2010, Renosh2015}. These methods ground on multiscale decompositions to provide non-linear characterizations of images across scales~\cite{Wendt2009,Arneodo2000, Gerasimova2014}. 

Signal and image processing methods based on Information theory~\cite{Shannon1948} were also developed during the last years~\cite{Bercher2000,RamirezReyes2016,Ma2018}. Information theory quantities used on multiscale decompositions were also studied in~\cite{Grazzini2002,Ahmed2012,Kim2021}. Notably~\cite{GraneroBelinchon2019,GraneroBelinchon2019a} provide a framework to characterize non-linear dependencies of stationary and non-stationary 1D-processes and \cite{GBelinchon2016, GraneroBelinchon2018} show the potentialities of this framework to study turbulent flows. Experimental signals of turbulent flows were also studied with information theory~\cite{Cerbus2013}. More recently, these methods have been used to study self-similar 2D-fields~\cite{Nicolis2020} and 3D turbulent data from direct numerical simulations~\cite{LozanoDuran2022}.

In this article we propose to generalize, for the description of 2D fields, the multiscale information theory framework developed in~\cite{GraneroBelinchon2018} for 1D processes. 
Our analysis characterizes
high order statistics across scales while being direction-dependent. First, we illustrate its 
use and behavior
on isotropic and anisotropic scale-invariant synthetic fields~\cite{Bonami2003,Bierme2009,Clausel2011}. We use, on the one hand two Gaussian fields: fractional Brownian motion (fBM) and anisotropic fractional Brownian motion (a-fBm), and on the other hand, two non-Gaussian ones: multifractal random walk (MRW) and anisotropic multifractal random walk (a-MRW). Our analyses recover the scale-invariant parameters of the fields as well as the anisotropy in the case of the a-fBm and the a-MRW.  

Then, to show a real case application of our methodology, we turn to fluid turbulence, where the main theory was issued for homogeneous and isotropic flows but all real life applications are notoriously anisotropic. We characterize two turbulent flows: a forced isotropic and homogeneous flow and a channel flow which is anisotropic due to the effect of the walls of the channel.
These flows are described by 3D velocity fields, $(u_x,u_y,u_z)$, and so comparing the results on the different components of velocity can illustrate a second kind of anisotropy~\cite{Kurien2000}. In both flows, the proposed approach is able to describe the scale-invariance of turbulence along each direction of analysis. Moreover in the channel flow, we observe an anisotropy in the energy distribution while the energy cascade and intermittency seems to remain isotropic. Indeed the anisotropy in the energy distribution is introduced by the mean velocity profile, whose shape depends on the configuration of the flow. With a proper substraction of the mean velocity profile requiring high-order multiscale decomposition~\cite{Cho2019, Angriman2022} we show that even in the channel flow configuration, the turbulent velocity fluctuations remain isotropic.

This work is structured as follows. Section~\ref{sec:theory} introduces the multiscale non-linear statistical framework: we present four quantities  for the characterization of 2D fields as well as their corresponding estimators. Section~\ref{sec:synthetic} presents our tests on isotropic and anisotropic scale-invariant synthetic fields and how their anisotropic multiscale properties are recovered. Section~\ref{sec:turb} describes our results on synthetic turbulent flows, while section~\ref{sec:Discussion} discusses how to eliminate the effect of the mean flow and recover the universal isotropy of the turbulent fluctuations.

\section{Multiscale characterization of anisotropic fields}
\label{sec:theory}

To assess the anisotropy of a two-dimensional field, especially in regards to its scale invariance properties, we 
estimate various higher-order statistical quantities of its increments. The first step is therefore a multiscale decomposition using spatial increments, presented in section~\ref{section:incrementsdef}, while the second step is the computation, after the decomposition, of four higher-order statistical quantities detailed in section~\ref{section:theory-stats}. The combination of the two steps leads to a multi-scale characterization of anisotropy, defined in~\ref{section:theory-scales}. The methodology used to obtain robust estimations is described in section~\ref{section:theory-metho}.

\subsection{Anisotropic multiscale decomposition}
\label{section:incrementsdef}
\begin{figure}[!t]
\centering
\includegraphics[width=0.7\linewidth]{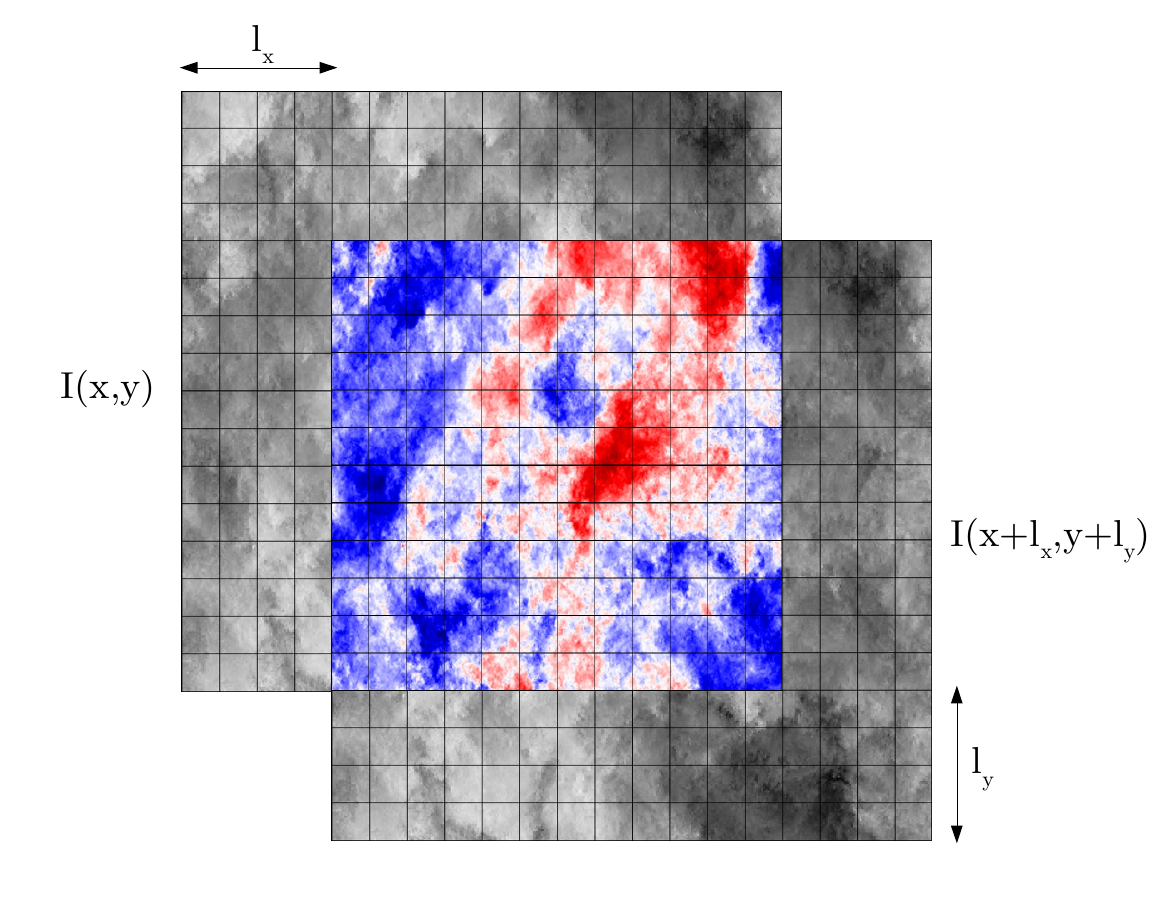}
\caption{\textbf{Increments of images:} Two-dimensional field $I(x,y)$ and its shifted version $I(x+l_x, y+l_y)$ are depicted using grey levels, while the corresponding increment $\delta_{l_x , l_y} I$, which is the difference of the two, is depicted in color.}
\label{fig:method}
\end{figure}

Given a two-dimensional field $I(x,y) \in \mathbb{R}$, where both $x$ and $y$ are defined in $\mathbb{R}$ and represent the coordinates of $I$, we propose a multiscale decomposition of $I(x,y)$ based on increments. Thus we define a two-dimensional scale $(l_x, l_y)$ and compute the increments~\cite{Kolmogorov1991, Frisch1995} over this scale as: 

\begin{equation}\label{eq:incrs}
\delta_{l_x , l_y} I =  I(x+l_x, y+l_y) - I(x, y)
\end{equation}
This procedure is represented in Figure~\ref{fig:method}. A change from cartesian to polar coordinates $(l_x,l_y) \rightarrow (r, \theta)$ allows an analysis across the scale magnitude $r$ and its direction $\theta$ defined from $\left(l_x,l_y\right)=\left(r \cos \left( \theta \right), r \sin \left( \theta \right) \right)$.

\subsection{Statistical description}
\label{section:theory-stats}

To characterize the probability density function (PDF) of a generic two dimensional random variable $U(x,y)$, we choose to focus on one side on the first two high-order statistical centered moments: the skewness which characterizes the asymmetry of the PDF and the kurtosis which characterizes the relative weight of its tails. On the other side, we choose two measures from information theory: the Shannon entropy which quantifies the total amount of information in $U$~\cite{Shannon1948}, and the Kullback-Leibler distance from Gaussianity, which quantifies how the PDF of $U$ deviates from a Gaussian PDF~\cite{GraneroBelinchon2018}.

Noting $\sigma^2_{U}$ the variance of the random variable $U$ and $p$ its PDF, we define:

\begin{eqnarray}
S(U) &=& \frac{1}{\sigma_{U}^{3}} \int_{\mathbb{R}} (u-\E[u])^3 p(u) du  \label{eq:Ske}\\ 
F(U) &=& \frac{1}{\sigma_{U}^{4}} \int_{\mathbb{R}} (u-\E[u])^4 p(u) du  \label{eq:Kur}\\ 
H(U) &=& \int_{\mathbb{R}} p(u) \log(p(u)) du  \label{eq:H}\\ 
D(U)&=& \int_{\mathbb{R}} p(u) \log\left(\frac{p(u)}{p_{G}(u)}\right) du \label{eq:K}
\end{eqnarray}

\noindent where $p_{G}$ is the Gaussian PDF with the variance $\sigma^2_{U}$ and a zero mean. 
None of these four quantities depends on $\E[u]$.

The entropy $H(U)$ characterizes the complexity of $U$; it depends on its PDF and hence on all statistical moments except the first order one. Generally, the main contribution to the entropy comes from the variance $\sigma^2_{U}$ while higher order moments are expected to have only a slight influence. In the case of a Gaussian variable $U_G$ its entropy is defined uniquely by its variance $\sigma^2_{U_G}$:

\begin{equation}
H(U_G) = \frac{1}{2} \log(2 \pi e \sigma^2_{U_G}) \,. \label{eq:HSG}
\end{equation}    

From eq.(\ref{eq:HSG}) we define $H_G(U)$, the entropy under Gaussian hypothesis of $U$, as the entropy of the Gaussian field with the same variance $\sigma^2_{U}$ as $U$:
\begin{equation}
H_G(U) = \frac{1}{2} \log(2 \pi e \sigma^2_{U}) \label{eq:HSG2} \,.
\end{equation} 

Hence, the distance from Gaussianity of $U$ can be expressed as the difference between its entropy under Gaussian hypothesis and its genuine entropy:

\begin{equation}
D(U)= H_G(U) - H(U) \geq 0 \label{eq:K2} \,.
\end{equation} 

The maximum entropy principle states that for a given variance the Gaussian PDF maximizes the entropy~\cite{Jaynes1957, Jaynes1957a}, hence the inequality in eq.(\ref{eq:K2}).

The distance from Gaussianity (\ref{eq:K}) or (\ref{eq:K2}) provides an insight on the signal different from the entropy~(\ref{eq:H}), since it focuses on moments of higher order and does not depend on the variance anymore. It can be used to describe the deformation of the PDF, just like the flatness does, but incorporating information from all other higher order moments~\cite{GraneroBelinchon2018}.

\subsection{Multiscale characterization of anisotropy}
\label{section:theory-scales}

We then define for the field $I$ the skewness, flatness, entropy and distance from Gaussianity at the scale $(l_x, l_y)$ by applying definitions (\ref{eq:Ske}), (\ref{eq:Kur}), (\ref{eq:H}) and (\ref{eq:K}) to the increment fields $U=\delta_{l_x , l_y} I$ defined in (\ref{eq:incrs}). We note these 4 quantities $S_{l_x,l_y}(I)$, $F_{l_x,l_y}(I)$, $H_{l_x,l_y}(I)$ and $D_{l_x,l_y}(I)$ respectively.  We then note $S_{r, \theta}(I)$, $F_{r,\theta}(I)$, $H_{r,\theta}(I)$ and $D_{r,\theta}(I)$ the same quantities as functions of polar coordinates.

By varying the two-dimensional scale $(l_x,l_y)$, or equivalently $(r,\theta)$, one is able to characterize the PDF of $I$ across scale magnitudes and directions. Then $\mathcal{S}_{l_x,l_y}(I)$ and $\mathcal{F}_{l_x,l_y}(I)$ describe how the asymmetry and tails of the PDF are evolving with the scale, while $H_{l_x,l_y}(I)$ and $D_{l_x,l_y}(I)$ characterize respectively the evolution of information and the deformation of the PDF across scale magnitudes and directions.

\subsection{Numerical implementation}\label{section:theory-metho}

In the following, we analyze images of typical size 8192 $\times$ 8192 pixels, unless noted otherwise. Images are here considered as individual realizations of stochastic fields superimposed or not over a mean behavior. Moreover, we consider that the fields are homogeneous, and so their statistical properties do not depend on the studied region. Estimation of the statistical moments are performed in Python 3.9 with non-biased estimators provided by SciPy library 1.7.3. Shannon entropy estimations are performed with our own implementation of Kozachenko and Leonenko $k$-nearest neighbors ($k$-nn) algorithm~\cite{Kozachenko1987}, which was shown to have a reduced bias compared to other Shannon entropy estimators~\cite{Kozachenko1987, Gao2018}.
The distance from Gaussianity is computed with the Kozachenko and Leonenko $k$-nn estimator for the first term $H(\delta_{l_x,l_y}I)$ in eq.(\ref{eq:K2}) and the SciPy unbiased estimator of the variance with eq.(\ref{eq:HSG}) for the second term $H_G(\delta_{l_x,l_y}I)$ in eq.(\ref{eq:K2}).

The only adjustable parameter in the $k$-nn algorithms is the number $k$ of neighbors, and we set it to $k=5$ for all estimations reported in this article~\cite{GraneroBelinchon2019a}.
The bias and variance of the corresponding estimators depend on $k$ and the number $N_{\text{eff}}$ of effective points over which the statistics are computed. In this article, we choose $N_{\text{eff}}=2^{12}$ points, which was proven satisfying by a detailed study~\cite{GraneroBelinchon2019a} for stationary long-range and short-range dependence processes with Gaussian and non-Gaussian statistics. For the values of $k$ and $N_{\text{eff}}$ chosen, the bias is negligible ($<0.005$).

We also follow a Theiler prescription~\cite{Theiler1986}: for a given scale $(l_x, l_y)$ or $(r, \theta)$, the $N_{\text{eff}}$ points must be sampled in the image in such a way that these points should all be distant one from another by at least $r_{\rm Theiler}\equiv r$. This prescription ensures that statistical dependencies on scales smaller than $r$ are not taken into account. Unfortunately, when the scale magnitude $r$ is large, the number of available points in a finite image may be smaller than the required $N_{\text{eff}}$. In that case, we alter the original prescription and require that points must be distant one from another by at least $r_{\rm Theiler}$=50 pixels. At this scale, spurious dependencies have already decreased and are fairly small. This trade-off allows us to explore scales $l_x$ and $l_y$ up to 640 pixels, while having enough points in the image to analyse 5 realizations. Each one of these realizations contains $N_{\text{eff}}$ points randomly sampled from the full domain of study, so they provide a global description. Despite the small impact of spurious correlations on the skewness and flatness, we use for the sake of simplicity this adapted Theiler prescription in the estimation of all the statistical quantities: $S$, $F$, $H$ and $D$. This procedure allows us to sample the whole image, perform estimations on multiple realizations and reduce computation time.

\section{Synthetic anisotropic scale-invariant stochastic fields}
\label{sec:synthetic}
\subsection{Scale invariance}

A field is {\bf scale-invariant} if the statistical moment of any order $q$ of its increments behaves as power law of the increment's size, i.e. in the case of one dimensional field $I(x)$ with increments $\delta_{l} I$:
 
\begin{equation}
\E \left(|\delta_{l} I|^q \right)\underset{l\rightarrow 0}\sim  k_q l^{\zeta(q)}
\end{equation}
with $k_{q}$ constants that depend on the order of the statistical moment and $\zeta(q)$ the scaling function which is concave.

We distinguish two main families of scale-invariant fields: monofractal and multifractal ones. 

{\bf Monofractal} fields are characterized by a linear scaling function $\zeta(q)=q{\cal H}$. The slope ${\cal H}$, which characterizes the roughness of the field, is called the Hurst exponent.

{\bf Multifractal} fields have a non-linear scaling function that can be approximated by $\zeta(q)=q{\cal H}-\frac{c_2}{2}q^2$, where ${\cal H}$ characterizes the most common roughness and $c_2>0$ is the intermittency coefficient characterizing how wide-ranging are the existing singularities. The nonlinearity of $\zeta(q)$ implies that the shape of the PDF of the increments evolves across scales and so the field is no more jointly Gaussian.

On the one hand for monofractal fields, the entropy of the increments behaves as the logarithm of the scales $H_{l}(I)\sim {\cal H} \log(l)$ and the flatness, skewness and Kullback-Leibler distance across scales are constants. Consequently, the shape of the PDF of the increments does not evolve across scales.
On the other hand for multifractal fields, the entropy of the increments still behaves mainly as $H_{l}(I)\sim {\cal H} \log(l)$ but the flatness is no more constant and decreases as a power law $F_{l}(I)\underset{{l \rightarrow 0}}\sim l ^{-4c_2}$. Thus, the Kullback-Leibler distance also decreases and goes to $0$ at large scale.

Finally, in the case of two dimensional fields $I(x,y)$ the moments of the increments depend on both the size $r$ and direction $\theta$ of the increments: 
\begin{equation}
\E \left(|\delta_{r,\theta} I|^q \right)\underset{r\rightarrow 0}\sim  k_{\theta,q} r^{\zeta_\theta(q)}
\end{equation}
The process is considered {\bf anisotropic scale-invariant} if its scaling function $\zeta_{\theta} (q)$ obtained from the statistics of the directional increments $\delta_{r,\theta} I$ depends on the angle $\theta$.

\subsection{Anisotropic scale-invariant random fields generation}

Homogeneous, isotropic and self-similar $d$-dimensional fields can be modelled with the following stochastic $d$-dimensional integral \cite{Robert2008, Pereira2016}:

\begin{equation}
I_{\mathcal{H},c_2,\eta}(\mathbf{z}) = \int_{\R^d} P_{\mathcal{H},\eta} 
( \mathbf{z}-\mathbf{z^\prime})
M_{c_2,\eta}(\mathbf{z^\prime})
W(\mathbf{z^\prime})
d\mathbf{z^\prime}
\label{def:field}
\end{equation}

\noindent where $\mathbf{z}=(z_1,z_2,\cdots,z_d)$ and $\mathbf{z^\prime}$ denote $d$-dimensional position vectors, $W(\mathbf{z^\prime})$ is a Gaussian white noise, $0<\mathcal{H}<1$ the Hurst exponent and $c_2$ the intermittency coefficient.

The first term, $P_{\mathcal{H},\eta}(\mathbf{z}) = \frac{1}{||\mathbf{z}||_\eta^{d/2-\mathcal{H}}}$, is a kernel providing a power spectrum with a power law behavior of exponent $2\mathcal{H}$. The norm $||\mathbf{z}||_\eta = \sqrt{||\mathbf{z}||^2+\eta^2}$ in the denominator is a regularized ${\cal L}^2$-norm that ensures the convergence of the integral when $0<\mathcal{H}<0.5$, with $\eta>0$ being the regularization scale and $||.||$ the ${\cal L}^2$-norm.
The second term, $M_{c_2,\eta}(\mathbf{z^\prime})$ is a  multiplicative chaos \cite{Robert2008, RhodesVargas2014} defined as:

\begin{equation}
M_{c_2,\eta}(\mathbf{z^\prime})=e^{-\sqrt{c_2} X_\eta(\mathbf{z^\prime})-c_2\E\{X_\eta^2(\mathbf{z^\prime})\}}
\end{equation}
\noindent where $X_\eta(\mathbf{z^\prime})$ is a log-correlated Gaussian noise with autocovariance function:

\begin{equation}\label{eq:autocov}
\E \{X_\eta(\mathbf{z})X_\eta(\mathbf{z^\prime}) \}  \underset{||\mathbf{z}-\mathbf{z^\prime}||_{\eta}\rightarrow 0}\sim -\log(||\mathbf{z}-\mathbf{z^\prime}||_\eta)
\end{equation}


When the intermittency coefficient $c_2=0$ in (\ref{def:field}) the stochastic process $I_{\mathcal{H},c_2=0,\eta}$ is a fractional Brownian motion of parameter $\mathcal{H}$. This process is a Gaussian non stationary process with stationary increments and is characterized by a linear scaling function $\zeta(q)=q\mathcal{H}$.  When $c_2>0$, the process is no more Gaussian and its scaling function behaves as $\zeta(q)=q(\mathcal{H}+c_2/2)-c_2q^2/2$. For one dimensional processes this definition corresponds to the Multifractal Random Walk first proposed in~\cite{Bacry2001}.

Anisotropy is added as in~\cite{Bonami2003,Bierme2009} by defining a transformation matrix $E$ satisfying $\mbox{Tr}(E)=d$ and replacing the $\mathcal{L}^2$-norm by a positive function $f$ satisfying the homogeneity relationship $f(a^{E}\mathbf{z})=af(\mathbf{z})$ on $\mathbb{R}^d$. Then, the kernel of (\ref{def:field}) reads: 

\begin{equation}
P_{\mathcal{H},E,\eta} (\mathbf{z})= \frac{1}{(f(\mathbf{z})+\eta)^{\frac{d}{2}-\mathcal{H}}}
\label{anikernel}
\end{equation}

In this study, we set $d=2$, we note the vector of position $\mathbf{z}=(x,y)$ and use the diagonal matrix $E =\begin{pmatrix}\alpha_0&0\\0&2-\alpha_0\end{pmatrix}$ with the pseudonorm

\begin{equation}
f(\mathbf{z})=\left(\vert x \vert^{2/\alpha_0} + \vert y \vert^ {2/(2-\alpha_0)}\right)^{1/2} \,,
\end{equation}
\noindent where $0<\alpha_0<2$ is the anisotropy coefficient.

When using the anisotropic kernel (\ref{anikernel}) in (\ref{def:field}) and the pseudonorm $f$ in (\ref{eq:autocov}), the process is scale invariant along the direction $(l_x,l_y)=(l^\alpha,l^{2-\alpha)})$ with the scaling characterized by 

\begin{equation}
\E \left( \left( \delta_{l^\alpha,l^{2-\alpha}} I_{\mathcal{H},E,c_2,\eta} \right)^q \right) \underset{l\rightarrow 0}\sim l^{\zeta_\alpha(q)}
\end{equation}
 
\noindent where $\zeta_\alpha(q)=(q(\mathcal{H}+c_2/2)-c_2q^2/2)\max\left(\frac{\alpha}{\alpha_0}, \frac{2-\alpha}{2-\alpha_0}\right)$ and $\alpha$ defines a direction of analysis. If $\alpha_0=1$, $E$ is the identity matrix, $f$ is the $\mathcal{L}^2$-norm and we recover the isotropic case. In the particular case of $c_2=0$ we recover the Operator Scaling Gaussian Random Field (OSGRF) studied in detail in~\cite{Clausel2011, Roux2013}.

The fields generated following the above definition depend on three parameters: the Hurst exponent $\mathcal{H}$, the multifractal coefficient $c_2$ and the anisotropy coefficient $\alpha_0$. In the following, we name the family of processes with $c_2 =0$ fractional Brownian motions and the family of processes with $c_2 \neq0$ multifractal random walks.

\subsection{Validation of multiscale characterization of anisotropy}

\begin{figure}[!ht]
\centering
\includegraphics[width=0.7\linewidth]{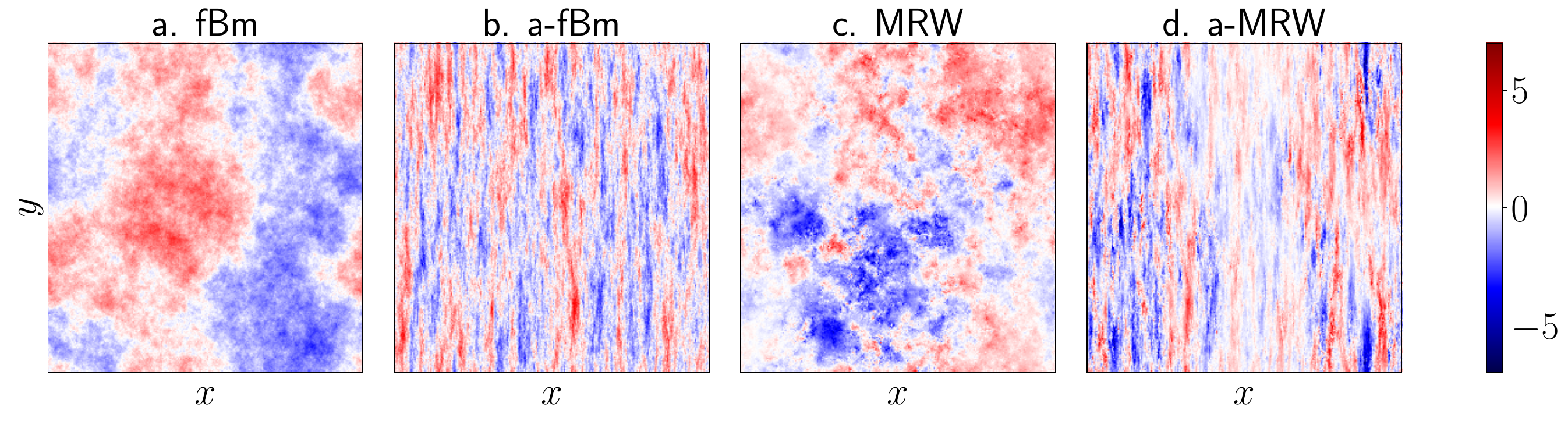}
\caption{{\bf Synthetic scale-invariant stochastic fields.} \quad Four examples of two-dimensional fields of size $N=8192^2$ with the same Hurst exponent $\mathcal{H}=\frac{1}{3}$. a) isotropic fBm, b) anisotropic fBm (both with $c_2=0$), c) isotropic MRW with $c_2=0.04$ and c) anisotropic MRW with $c_2=0.04$. The anisotropy coefficient of a-fBm and a-MRW is $\alpha_0=0.8$.}
\label{fig:mrw_Image}
\end{figure}

We first examine and compare two isotropic and two anisotropic scale invariant fields with $\mathcal{H}=1/3$. All fields are of size $N=8192^2$. The isotropic fields are a fractional Brownian motion and a Multifractal Random Walk; the fBm is monofractal ($c_2=0$) and the MRW has a multifractal parameter $c_2=0.04$. The anisotropic fields are built as described in the previous section to be  anisotropic versions of these two fields: an anisotropic fBm and an anisotropic MRW, both with the anisotropy coefficient $\alpha=0.8$ and with the same $c_2$ as their isotropic counterpart. The four fields are presented in Figure~\ref{fig:mrw_Image}. 

\begin{figure}[!ht]
\centering
\includegraphics[width=0.7\linewidth]{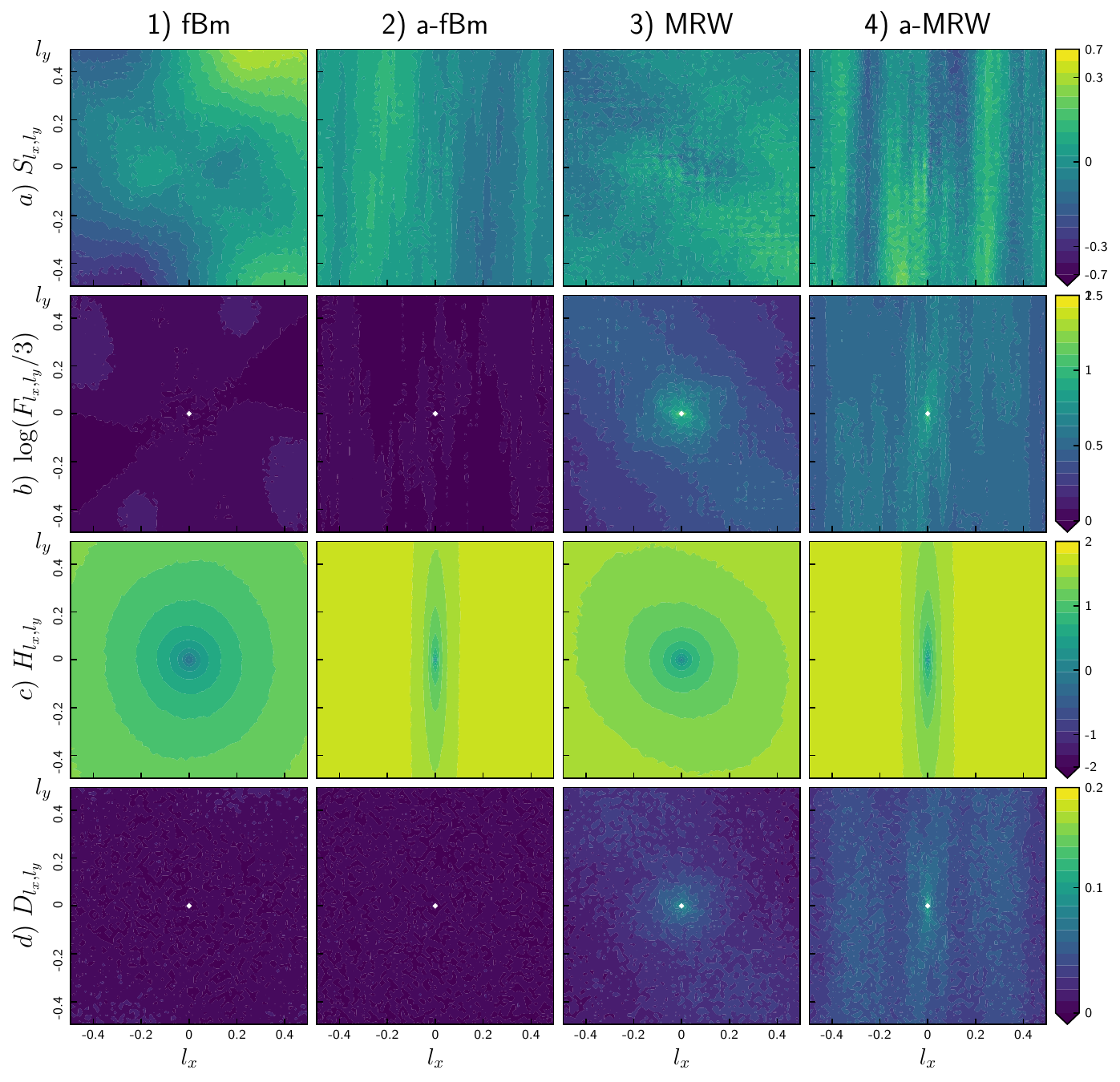}
\caption{{\bf 2D characterization of synthetic scale-invariant fields.} \quad Skewness, flatness, entropy and distance from Gaussianity across scales ($S_{l_x,l_y}$, $\log(F_{l_x,l_y}/3)$, $H_{l_x,l_y}$ and $D_{l_x,l_y}$), of a fBm, a-fBm, MRW and a-MRW, all with $\mathcal{H}=1/3$ and size $N=8192^2$. The anisotropy coefficient of both anisotropic fields is $\alpha_0=0.8$ and the multifractal parameter of both MRWs is $c_2=0.04$.}
\label{fig:mrw_Cartesian2d}
\end{figure}

\begin{figure*}[!ht]
\centering
\includegraphics[width=0.7\textwidth]{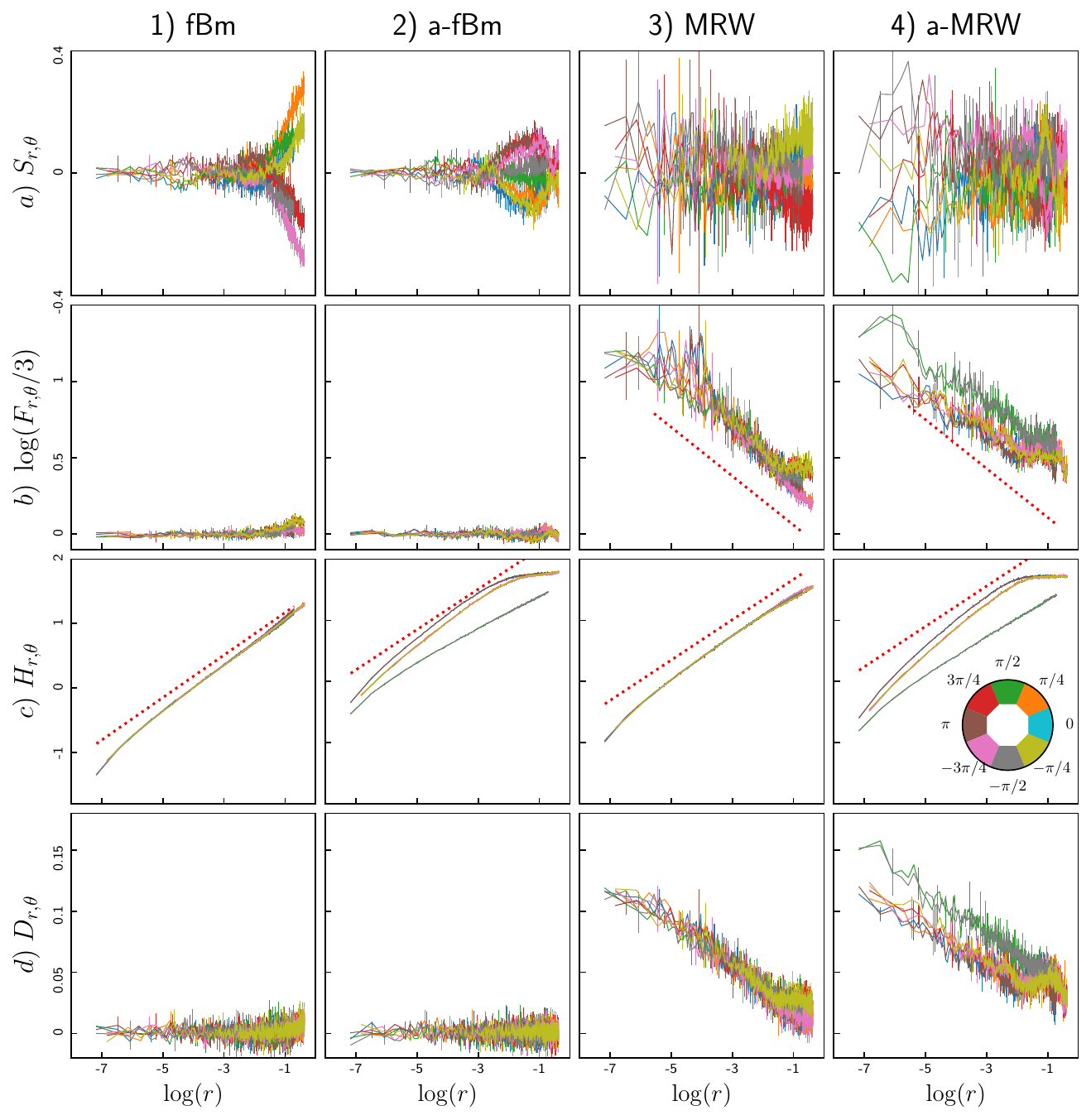}
\caption{{\bf 1D $\theta$-transects characterization of synthetic scale-invariant fields.} \quad Skewness, flatness, entropy and distance from Gaussianity across scales ($S_{r,\theta}$, $\log(F_{r,\theta}/3)$, $H_{r,\theta}$ and $D_{r,\theta}$), of a fBm, a-fBm, MRW and a-MRW, all with $\mathcal{H}=1/3$ and size $N=8192^2$. The MRWs have $c_2=0.04$ and the anisotropic fields have $\alpha_0=0.8$. Red dotted lines in c) illustrate a slope of $\mathcal{H}$ while red dotted lines in b) illustrate a $-4 c_2$ slope. For each synthetic process eight different directions are studied : $\theta\in\{-3\pi/4, -\pi/2, -\pi/4, 0, \pi/4, \pi/2, 3\pi/4, \pi\}$.}
\label{fig:mrw_H-K}
\end{figure*}

Figure~\ref{fig:mrw_Cartesian2d} shows how $S_{l_x,l_y}$, $\log(F_{l_x,l_y}/3)$, $H_{l_x,l_y}$ and $D_{l_x,l_y}$ evolve with the scale in cartesian coordinates for the four studied fields. To better examine and quantify the isotropy properties of the fields, we use polar coordinates and represent in Figure~\ref{fig:mrw_H-K}  $S_{r,\theta}$, $\log(F_{r,\theta}/3)$, $H_{r,\theta}$, and $D_{r,\theta}$ as functions of $\log(r)$ for eight directions $\theta\in\{-3\pi/4, -\pi/2, -\pi/4, 0, \pi/4, \pi/2, 3\pi/4, \pi\}$. Errorbars indicate the standard deviation across realizations.

The skewness of all these fields fluctuates around zero, which indicates that the PDFs of their increments remain symmetrical for all scales. We also recover that the skewness is an odd function of the increment: $S_{-l_x,-l_y} = -S_{l_x,l_y}$, or $S_{r,\theta+\pi} = -S_{r,\theta}$. For the fBm however, although increments are theoretically Gaussian and hence $S_{l_x,l_y}=0$ at all scales, we observe that the skewness increases for larger scales; this is due to finite-size effects which, curiously, only impact marginally the other statistics. The skewness of a-fBm, MRW and a-MRW is less impacted by these effects.

The entropy $H$ evolves from small values at small scales to large values at large scales, which indicates an increase of the complexity when increasing the scale. In Figure~\ref{fig:mrw_H-K} c), we see that this increase is linear in $\log(r)$ for all the fields. In the case of a-fBm and a-MRW, Figures~\ref{fig:mrw_Cartesian2d} b) and d) show that iso-entropy lines are thin vertical ellipses instead of circles as for the fBm and MRW: the entropy exhibits a very strong anisotropic behavior with a very fast increase from small to large scales along the $l_x$ direction and a much slower increase in the $l_y$ direction. From Figure~\ref{fig:mrw_H-K} c), we find that for the isotropic fBM and MRW the entropy increases with a slope close to $\mathcal{H}$ independant of the direction $\theta$, as expected, while for the anisotropic fields the slope strongly depends on the direction of analysis.

Both the flatness $\log(F/3)$ and the distance from Gaussianity $D$ illustrate the monofractal nature of the fBms and the multifractal nature of MRWs. For the fBm and a-fBm, we see in Figures~\ref{fig:mrw_Cartesian2d} b) and d) that these two quantities vanishes everywhere; this is confirmed in Figure~\ref{fig:mrw_H-K}: $\log(F_{r,\theta}/3)$ and $D_{r,\theta}$ are zero for all scales magnitude $r$, as expected for a process with Gaussian statistics. On the contrary, for both the MRW and a-MRW, the flatness $\log(F_{r,\theta}/3)$ and the distance from Gaussianity $D_{r,\theta}$ decrease linearly when $\log(r)$ increases: this indicates that the statistics of the increments of both MRWs are non-Gaussian at smaller scales, and become closer to those of a Gaussian at larger scales. Moreover, we are able to distinguish between the MRW and a-MRW: for the MRW on one hand, $\log(F_{r,\theta}/3)\sim -4c_2\log(r) \, \, \forall \, \theta$ and so the Flatness provides a measure of $c_2$. For the a-MRW on the other hand, the slope of $\log(F_{r,\theta}/3)$ in $\log(r)$ depends on $\theta$ and can be smaller or larger than $-4c_2$ depending on the direction of analysis $\theta$.
The anisotropy of the a-MRW is thus quantifiable using the entropy, the flatness or the distance from Gaussianity. 

\section{Application to synthetic velocity fields in fluid turbulence}
\label{sec:turb}

Fluid turbulence is a 3-dimensional multiscale process where the energy cascades from larger scales down to smaller scales~\cite{Richardson1921}. The existence of this energy cascade implies a negative skewness for the velocity increments along the longitudinal direction of the flow~\cite{Kolmogorov1991,Frisch1995}. Moreover, turbulence exhibits intermittency: the probability distributions of the increments deform from Gaussian at large scales to non-Gaussian at small scales~\cite{Kolmogorov1962, Obukhov1962}. Characterizing intermittency thus requires the examination of high order statistics~\cite{Frisch1985,Frisch1995}. 

Within the classical framework of fully developed turbulence, the velocity field and its increments are statistically homogeneous and isotropic: all velocity components have identical statistics, and these do not depend on the direction but only on the magnitude of the scale. In a realistic configuration however, the velocity field is expected to be anisotropic; this is for example the case when the flow is restricted in a finite domain with a specific shape, or a finite domain with no-slip conditions at some of its boundaries. Nevertheless, if the Reynolds number is large enough, fluctuations of the velocity field are expected to be isotropic in the bulk. Indeed, it has been shown that they are always well described within this classical framework~\cite{,Frisch1995}.

\begin{figure}[!ht]
\centering
\includegraphics[width=0.7\linewidth]{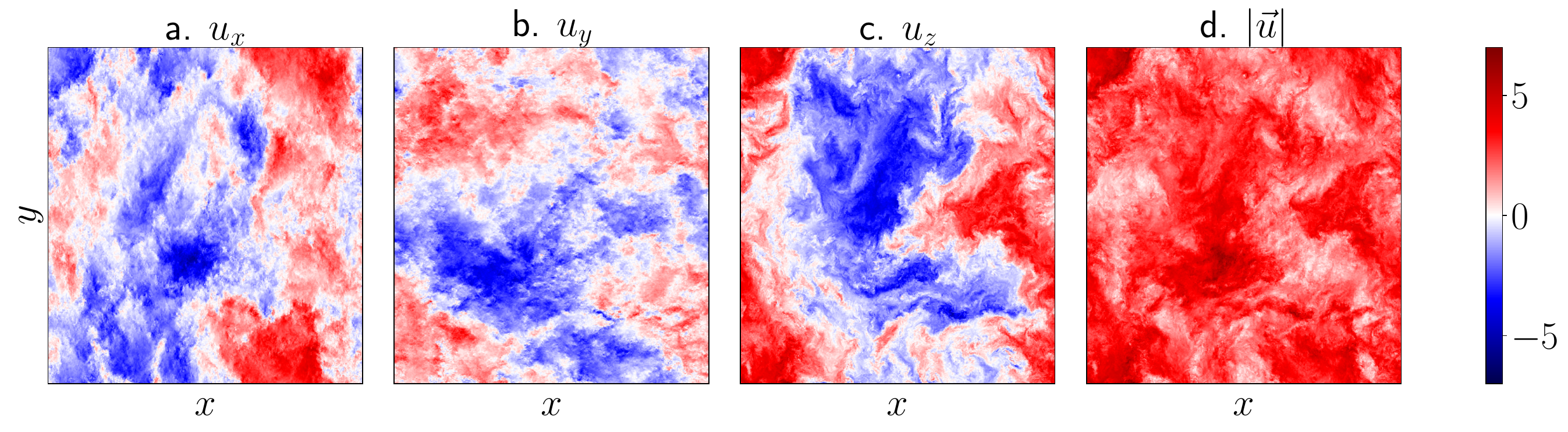}
\caption{ {\bf 2D transect of velocity fields from 3D forced isotropic turbulence.} \quad Two-dimensional transect from Johns Hopkins University Direct Numerical Simulation (http://turbulen\-ce.pha.jhu.edu). The $(x,y)$ coordinates span the whole spatial domain $[0, 2\pi]\times[0,2\pi]$; and $z=0.7853$ is fixed.
a) $u_x$, b) $u_y$, c) $u_z$ and d) $|\vec{u}|$.}
\label{fig:IsotropicJHU8192_Image}
\end{figure}

We study here two 2D transects of three dimensional turbulent velocity fields obtained by direct numerical simulation (DNS). For each transect, we analyze four images: one for each component $u_x$, $u_y$ and $u_z$ of the original 3-d velocity field, and one for its modulus $|u|$. This allows us to study not only the isotropy of each image, by tracking the dependences on the direction of analysis $\theta$, but also the isotropy of the velocity vector field by comparing our measurements from one image to another, {\em i.e.}, depending on the velocity component.

The first transect is from a DNS of forced isotropic turbulence on a periodic cubic grid $[0; 2\pi]^3$ with $8192^3$ points~\cite{Yeung2015}. The Reynolds number is $R_e=1256.8$, so the flow can be considered as exhibiting fully developed turbulence; the integral scale is $L=1.2438$ and the Kolmogorov scale is $\eta=5.89 \times 10^{-3}$ in simulation units. The transect we chose is defined by $z=0.7853$, see Figure~\ref{fig:IsotropicJHU8192_Image}. Images are 8192$\times$8192 pixels, with the pixel width being $7 \times 10^{-4}$ simulation units. 

\begin{figure}[!ht]
\centering
\includegraphics[width=0.7\linewidth]{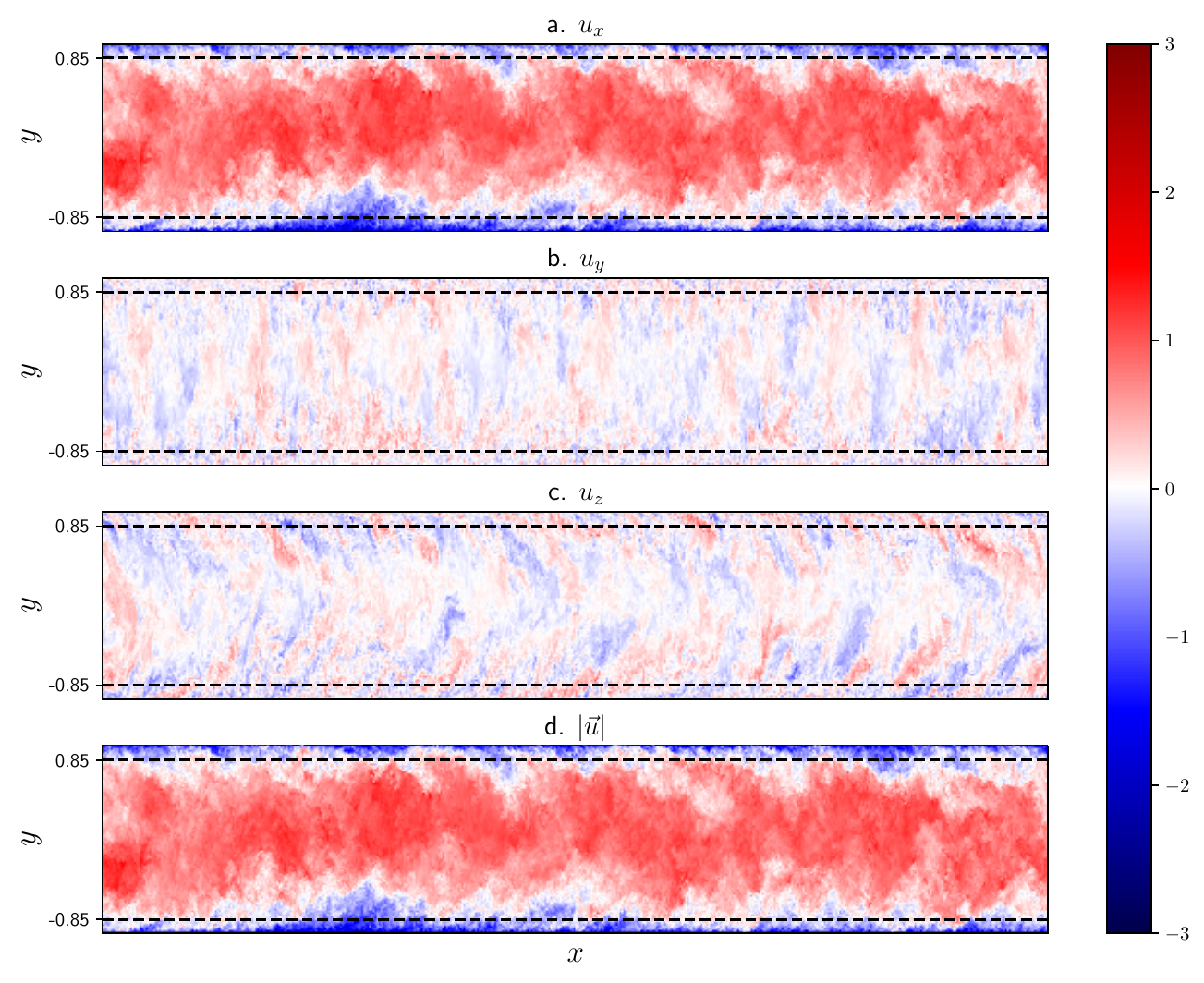}
\caption{ {\bf 2D transect of velocity fields from 3D anisotropic turbulence.} \quad Two-dimensional transect of the 3-d velocity field from JHU Channel turbulent flow Direct Numerical Simulation (http://turbulen\-ce.pha.jhu.edu). 
The $(x,y)$ coordinates span the full spatial region $([0, 8\pi], [-1, 1])$ and $z=\frac{3\pi}{2}$ fixed.
a) $u_x$, b) $u_y$, c) $u_z$ and d) $|\vec{u}|$. The horizontal black dashed lines indicate the values $y\in\{-0.85,0.85\}$.}
\label{fig:ChannelJHU_Image}
\end{figure}

\begin{figure}[!ht] 
\includegraphics[width=0.8\textwidth]{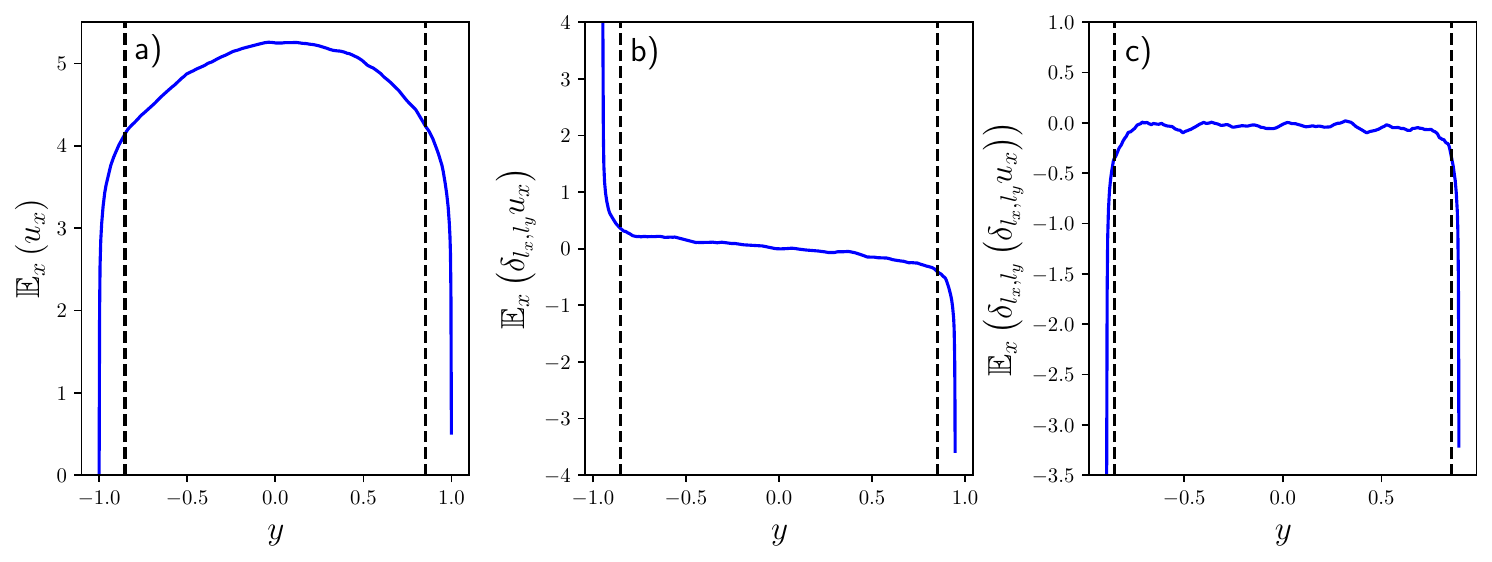}
	\caption{\textbf{Anisotropic turbulent velocity profiles.} Mean profile of a) velocity component $u_x$, b) increment $\delta_{lx,ly} u_x$, c) second order increment $\delta_{lx,ly} \left( \delta_{lx,ly} u_x\right)$ as functions of $y$. These profiles have been obtained by averaging over $x$ for a fixed $y$. Increments are defined on scale $l_x=0$ and $l_y=0.104$. The black vertical dashed lines indicate the central region $y \in[-0.85,0.85]$ used in the analysis, away from boundary layers.}
	\label{fig:profiles}
\end{figure}

The second transect is from a DNS of a turbulent channel flow with periodic boundary conditions in the $x$ and $z$ directions, and no-slip conditions at the top and bottom walls ($y$ direction)~\cite{Lee2015}. This flow is strongly non-isotropic and inhomogeneous and so it is a good case study for our method. The friction-velocity Reynolds number of the flow is $R_{\tau}=5200$, and we consider it a fully developed turbulent flow. The domain of the simulation is $(x,y,z) \in [0, 8\pi[ \times [-1, +1] \times [0, +3\pi[$. We study the 2D transect at $z=\frac{3\pi}{2}$; which gives us images of size $10240 \times 1536$ pixels in the $x$ and $y$ directions; here a pixel corresponds to $\pi/1280$ along $x$ and non-regular sampling along $y$, see Figure~\ref{fig:ChannelJHU_Image}. The anisotropy of the velocity field is two-fold. First, there is no mean flow in the $y$ and $z$ directions while there is a mean velocity along the $x$ direction. Second, the mean velocity profile $\mathbb{E}_{x}\left(u_x\right)(y)$ is inhomogenous due to the existence of boundary layers around $y=\pm 1$. As illustrated in Figure ~\ref{fig:profiles} a), where the streamwise component of the velocity is depicted, the flow has an approximate parabolic mean velocity profile far from the boundaries. In this work, we restrict our analysis to the bulk domain $y \in [-0.85, 0.85]$, in order to discard the strongest gradients close to the walls.

\subsection{Forced isotropic turbulence}

\begin{figure}[!ht] 
\includegraphics[width=0.7\linewidth]{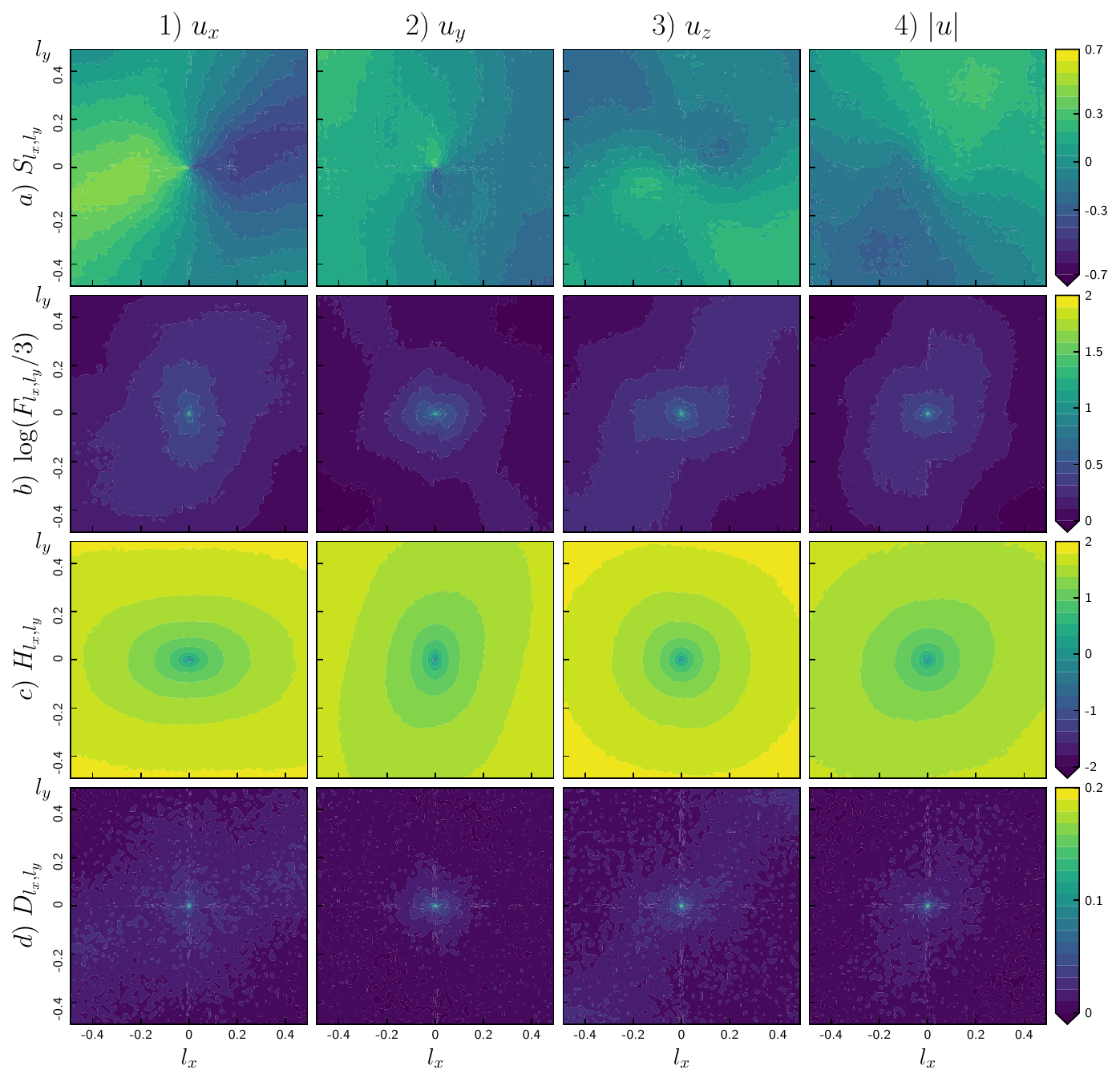}
	\caption{ {\bf 2D characterization of velocity fields of forced isotropic turbulence.} \quad Skewness, flatness, entropy and distance from Gaussianity across scales ($S_{l_x,l_y}$, $\log(F_{l_x,l_y}/3)$, $H_{l_x,l_y}$ and $D_{l_x,l_y}$), of the velocity fields  $u_x$, $u_y$, $u_z$ and $|\vec{u}|$.}
	\label{fig:IsotropicJHU8192_Cartesian2d_ux_uy_uz_abs}
\end{figure}

\begin{figure}[!ht]
\centering
\includegraphics[width=0.7\linewidth]{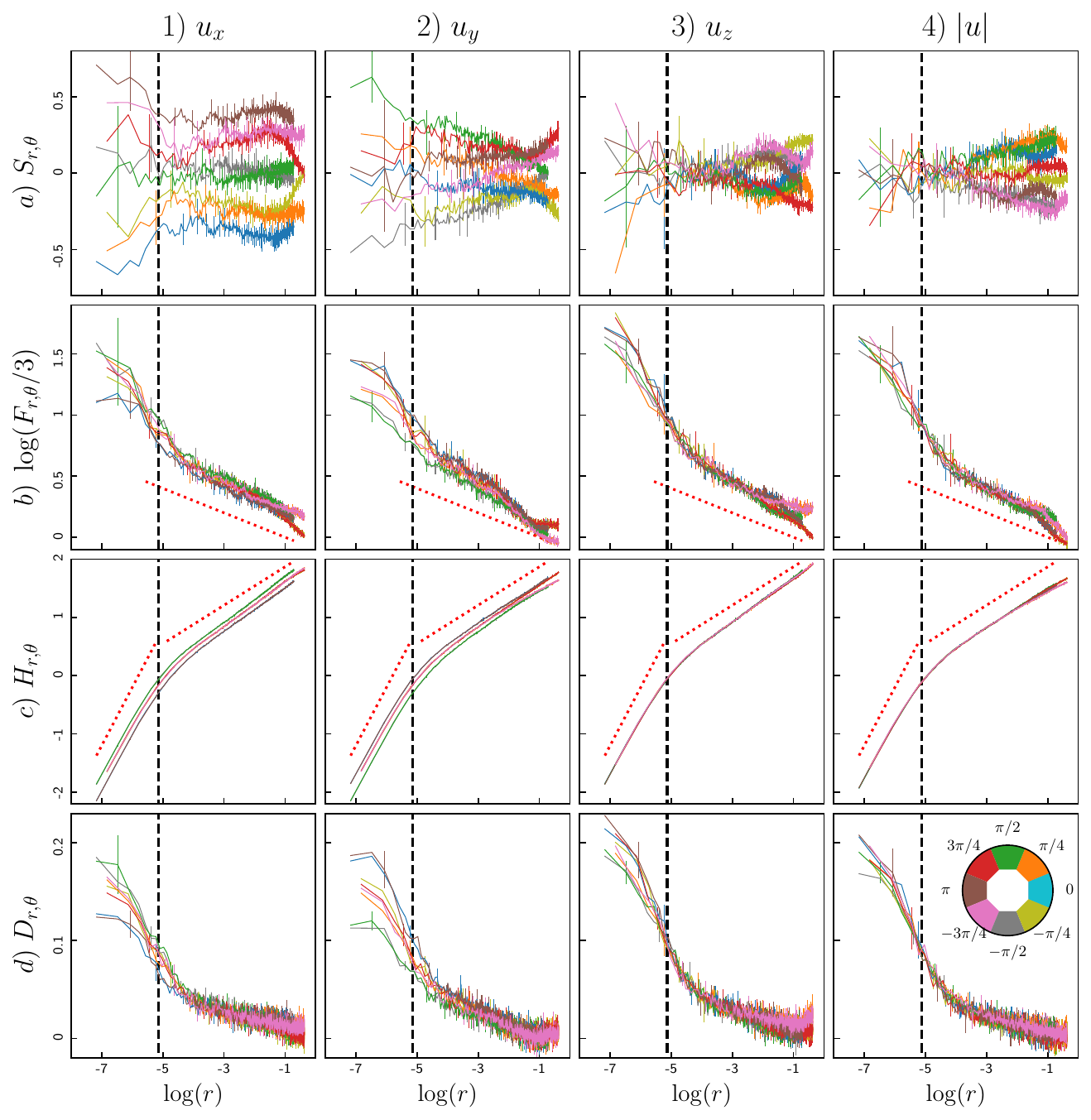}
\caption{{\bf 1D $\theta$-transects characterization of velocity fields of forced isotropic turbulence.} \quad a) Skewness, b) flatness, c) entropy and d) distance from Gaussianity across scales ($S_{r,\theta}$, $\log(F_{r,\theta}/3)$, $H_{r,\theta}$ and $D_{r,\theta}$) of $u_x$, $u_y$, $u_z$ and $|\vec{u}|$. Errorbars indicate the standard deviation of the estimations over realizations. Red dotted lines in the entropy across scales illustrate the expected slope of $1$ in the dissipative domain and $1/3$ in the inertial one. Red dotted lines in the flatness plot illustrate the $-4 \, \times \, 0.025$ slope. For each velocity component 8 different directions are studied :  $\theta\in\{-3\pi/4, -\pi/2, -\pi/4, 0, \pi/4, \pi/2, 3\pi/4, \pi\}$.}
\label{fig:IsotropicJHU8192_H-K}
\end{figure}

Figure~\ref{fig:IsotropicJHU8192_Cartesian2d_ux_uy_uz_abs} shows how our four quantities $S_{l_x,l_y}$, $\log(F_{l_x,l_y}/3)$, $H_{l_x,l_y}$ and $D_{l_x,l_y}$ vary with the scales $(l_x,l_y)$, when applied to the four images corresponding to the three components and the modulus of the velocity field. Figure~\ref{fig:IsotropicJHU8192_H-K} presents a different view using polar coordinates, as a function of $\log(r)$ for eight equally spaced directions  $\theta\in\{-3\pi/4, -\pi/2, -\pi/4, 0, \pi/4, \pi/2, 3\pi/4, \pi\}$. 

The skewness $S_{r,\theta}$ evolves differently depending on the velocity component and the direction of analysis. Using the directions $x$ and $y$ of the initial velocity field transect, we are able to probe the longitudinal increments of the velocity components $u_x$ and $u_y$ respectively: these correspond to the angles $\theta=0$ and $\pi$ for the image of $u_x$, and the angles $\theta=\pm \pi/2$ for the image of $u_y$. For these specific directions, the 4/5th law of Kolmogorov~\cite{Frisch1995} predicts an almost constant skewness in the inertial domain. Indeed, we find for $u_x$ and $u_y$ that the skewness in the inertial domain is almost constant and maximum for the angles corresponding to the longitudinal direction, while it seems to increase in the dissipative domain. We are not able to probe the integral domain here, and cannot confirm that the skewness would then vanish. We moreover recover that the skewness is an odd function of the increment $S_{r,\theta} = -S_{r,\theta+\pi}$, see Figure~\ref{fig:IsotropicJHU8192_Cartesian2d_ux_uy_uz_abs}. In particular we observe this symmetry by looking at the longitudinal increments: for the component $u_x$ the skewness along the angles $\theta=0$ and $\theta=\pi$, which probe respectively increments $(l_x, 0)$ and $(-l_x, 0)$, just changes sign. For the component $u_y$, we also have that $S_{r,-\pi/2} = -S_{r,+\pi/2}$ for all $r$, using then the longitudinal direction along $y$. On the other hand, for $u_x$ and $u_y$ along transverse directions as well as for $u_z$ and $|u|$, the skewness oscillates around zero. These results illustrate that the skewness of each velocity component obeys the Kolmogorov theory along its longitudinal direction and shows no skewness in transverse ones. In the case of $|\vec{u}|$ the skewness is close to zero with a small increase at large scales maybe due to an asymmetrical compensation in the modulus. Moreover statistical effects can appear here because we only study a transect $z=0.7853$ of the full simulation cube, which severely restricts the spatial domain of analysis.

The entropy $H_{r,\theta}$ of the velocity components and modulus increases with the scale: the amount of information needed to characterize larger scales is greater than for smaller ones and the complexity of the velocity PDFs increases with the scale. As seen in Figure~\ref{fig:IsotropicJHU8192_H-K} c), this increase is linear in $\log(r)$ with a slope 1 in the dissipative domain and a slope $1/3$ in the inertial range, as expected. The anisotropy of the entropy can be seen in Figure~\ref{fig:IsotropicJHU8192_Cartesian2d_ux_uy_uz_abs} c): for the components $u_x$ and $u_y$, iso-entropy lines are ellipses, while they are almost circles for $u_z$ and $|\vec{u}|$. This is corroborated by looking at the dependence on the angle $\theta$ in Figure~\ref{fig:IsotropicJHU8192_H-K} c): for $u_x$ and $u_y$, $H_{r,\theta}$ depends slightly on the angle while its slope as a function of $\log(r)$ does not. For $u_z$ and $|\vec{u}|$, no clear dependence on $\theta$ is observed. We conclude that the entropy is almost isotropic for all components and modulus of the velocity field.

The flatness $F_{r,\theta}$, as well as the distance $D_{r,\theta}$ from Gaussianity, decrease as the scale $r$ increases and both vanish at large scales, for all velocity components and modulus, see Figures~\ref{fig:IsotropicJHU8192_Cartesian2d_ux_uy_uz_abs} b) and d) and Figures~\ref{fig:IsotropicJHU8192_H-K} b) and d). This indicates an evolution of the PDF of the increments of the velocity components and modulus from non-Gaussian at small scales to Gaussian at large scales, which characterizes the intermittency of the turbulent velocity field. Furthermore, we quantify that the deformation of PDF, as measured by $\log(F_{r,\theta}/3)$ and $D_{r,\theta}$, is stronger in the dissipative domain and weaker once in the inertial domain~\cite{Chevillard2005}. Quantitatively, $\log(F_{r,\theta}/3)$ decreases linearly with $\log(r)$ in the inertial domain with a slope $-4 c_2$ with $c_2=0.025$~\cite{Chevillard2012}. While some slight dependence on the angle $\theta$ can exist in the dissipative and integral domains, we conclude that both $F_{r,\theta}$ and $D_{r,\theta}(S)$ have an isotropic behavior in the inertial domain.

Because on the weak dependence of our results for a given velocity component on the direction $\theta$, and the weak variation from one component to another, which can be attributed to the choice of the transect and should average to zero when considering an ensemble of transects, we conclude that the velocity vector field from this DNS is isotropic.

\subsection{Turbulent channel flow}

\begin{figure}[!ht] 
\includegraphics[width=0.7\textwidth]{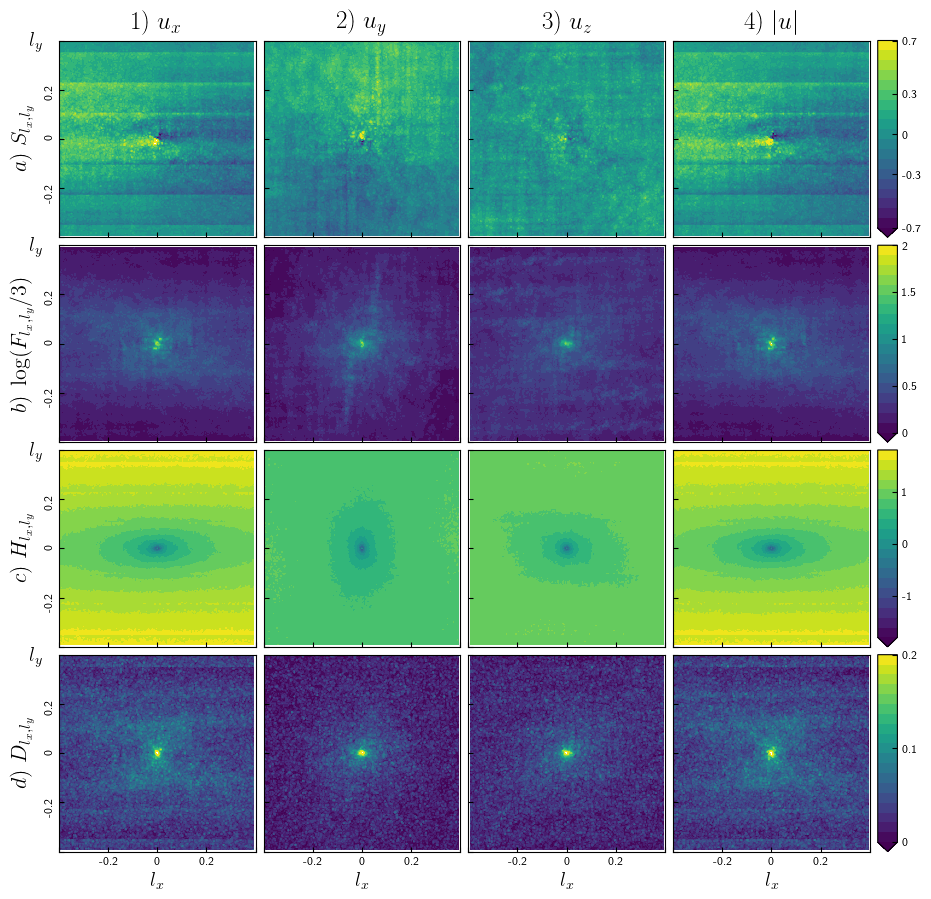}
	\caption{ {\bf 2D characterization of velocity fields of anisotropic turbulence.} \quad Skewness, flatness, entropy and distance from Gaussianity across scales ($S_{l_x,l_y}$, $\log(F_{l_x,l_y}/3)$, $H_{l_x,l_y}$ and $D_{l_x,l_y}$), of the velocity fields of the channel turbulent flow $u_x$, $u_y$, $u_z$ and $|\vec{u}|$.}
	\label{fig:ChannelJHU_Cartesian2d}
\end{figure}

Figure~\ref{fig:ChannelJHU_Cartesian2d} illustrates the evolution of $S_{l_x,l_y}$, $\log(F_{l_x,l_y}/3)$, $H_{l_x,l_y}$ and $D_{l_x,l_y}$ of $u_x$, $u_y$, $u_z$ and $|\vec{u}|$ along different directions and across scales. 
The skewness presents the same symmetries as before and also oscillates around zero. For $u_x$ and $|\vec{u}|$ strange horizontal lines appear along the $x$ axis which can be due to remaining effects of the walls.
As before the entropy across scales increases when the scale increases indicating an increase of complexity with the scale of analysis. The entropy of $u_x$ and $|\vec{u}|$ shows circular patterns at small scales which clearly become stretched ellipses at large scale. This illustrates the anisotropy of the flow. On the other hand for $u_y$ and $u_z$ the entropy remains isotropic. Finally, both $\log(F_{l_x,l_y}/3)$ and $D_{l_x,l_y}$ show similar patterns with high values at small scales that goes to $0$ at large scales, showing that the flow statistics then become Gaussian. The evolution of $\log(F_{l_x,l_y}/3)$ and $D_{l_x,l_y}$ seems isotropic for all the studied fields.

\begin{figure}[!ht]
\centering
\includegraphics[width=0.7\textwidth]{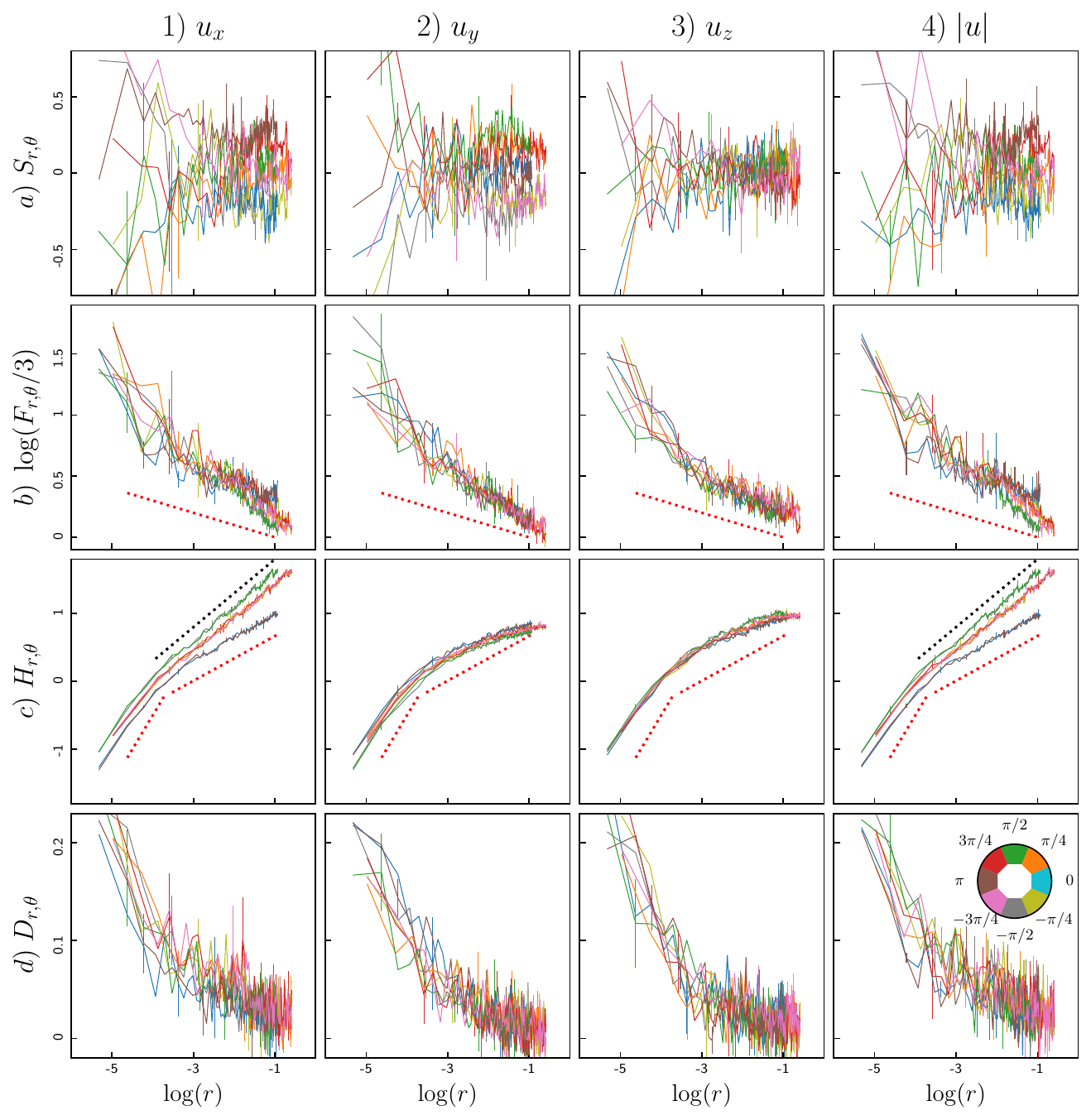}
\caption{{\bf 1D $\theta$-transects characterization of velocity fields of anisotropic turbulence.}  \quad Skewness, flatness, entropy and distance from Gaussianity across scales ($S_{r,\theta}$, $F_{r,\theta}$, $H_{r,\theta}$ and $D_{r,\theta}$) of $u_x$, $u_y$, $u_z$ and $|\vec{u}|$. Errorbars indicate the standard deviation of the estimations over realizations. Red dotted lines in the entropy across scales illustrate the expected slope of $1$ in the dissipative domain and $1/3$ in the inertial one. Black dotted line in the entropy across scales has a slope of $0.5$. Red dotted line in the flatness plot illustrate the $-4 \, \times \, 0.025$ slope. For each velocity component eight different directions are studied : $\theta\in\{-3\pi/4, -\pi/2, -\pi/4, 0, \pi/4, \pi/2, 3\pi/4, \pi\}$. A strong dependence in $\theta$ is observed in the behavior of $H_{r,\theta}$ for $u_x$ and $|\vec{u}|$, figures 1,c) and 4,c).}
\label{fig:ChannelJHU_H-K}
\end{figure}

Figure~\ref{fig:ChannelJHU_H-K} illustrates the evolution across scales $\log(r)$ of $S_{r,\theta}$ (a), $\log(F_{r,\theta}/3)$ (b), $H_{r,\theta}$ (c) and $D_{r,\theta}$ (d), and for $u_x$, $u_y$, $u_z$ and $|\vec{u}|$ along eight directions $\theta \in \{-3\pi/4, -\pi/2, -\pi/4, 0, \pi/4, \pi/2, 3\pi/4, \pi\}$. This figure supports the observations done for Figure~\ref{fig:ChannelJHU_Cartesian2d}. The skewness shows small variations around zero with slightly larger values at small scales as expected from theory. The flatness and the distance from Gaussianity across scales goes from Gaussian values at large scales to non-Gaussian ones at small scales. This evolution is the same independently of the direction of analysis and the studied field. So, intermittency seems to be isotropic in the studied domain. Moreover, it illustrates the behavior of $F_{r,\theta}/3$ in $r^{-4 \times 0.025}$ in the inertial domain. The entropy $H_{r,\theta}$ of $u_y$ and $u_z$ is isotropic and presents a slope of $1$ in the dissipative domain and $1/3$ in the inertial one independently of the direction $\theta$ of analysis. On the contrary, the entropy of $u_x$ and $|\vec{u}|$ are clearly anisotropic. Whereas in the dissipative domain the entropy behaves linearly with a slope of $1$ independently of $\theta$, in the inertial domain the directions parallel to the walls $\theta=\{0,\pi\}$ still present a slope of $1/3$ while the others exhibit a stepper behavior with a slope of $1/2$ in the directions perpendicular to the walls. These deviations from the Kolmogorov theory of homogeneous and isotropic turbulence~\cite{GraneroBelinchon2018} along the $\theta=\pm\pi/2$ directions are explained by the presence of the walls, which break the translational symmetry along the $y$ direction and introduce anisotropy.

\section{Discussion on the isotropy of turbulent fluctuations}
\label{sec:Discussion}

The anisotropic behavior of the entropy across scales of $u_x$ from the channel flow results from the existence of a non-linear mean velocity profile of $u_x$ along the $y$ direction, presented in Figure~\ref{fig:profiles} a). As a consequence, the mean profile of the increments of $u_x$ along $y$ still presents a linear trend, see Figure~\ref{fig:profiles} b), which leads to the anisotropy of the entropy. However, no trend is observed in the mean profile of the second order increments, defined as $\delta_{lx,ly} \left( \delta_{lx,ly} u_x\right)$, as can be seen Figure~\ref{fig:profiles} c).

\begin{figure}[!ht] 
\includegraphics[width=0.7\textwidth]{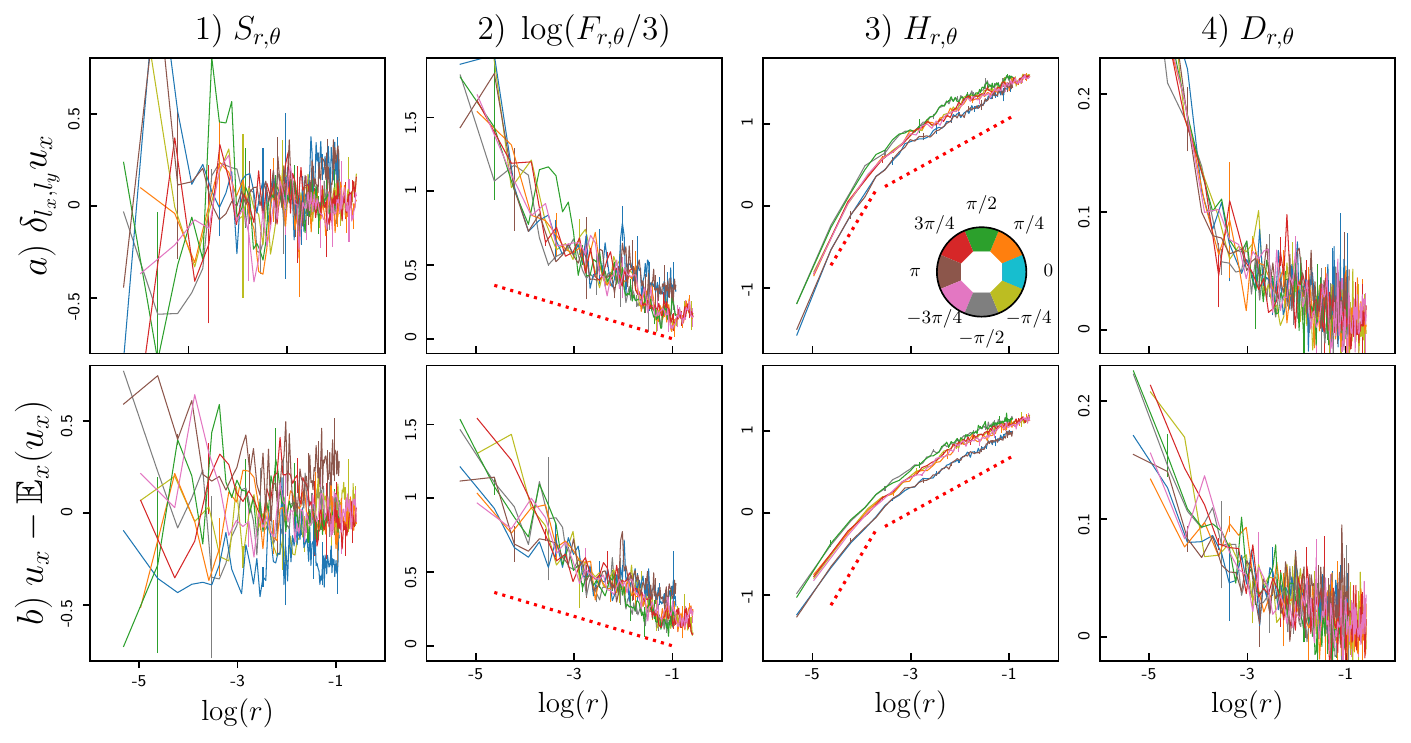}
\caption{\textbf{1D $\theta$-transects characterization of velocity fluctuations for anisotropic turbulence.} Skewness, flatness, entropy and distance from Gaussianity of a) second order increments of $u_x$ and b) increments of the velocity field without its mean behavior along the $y$ direction $u_x-\mathbb{E}_{x}(u_x)$.}
	\label{fig:2ndorder}
\end{figure}

Figure~\ref{fig:2ndorder} a) shows the results obtained when analysing the second order increments and Figure~\ref{fig:2ndorder} b) those obtained when considering the velocity field without its mean profile, $\tilde{u}(x,y)=u_x(x,y)-\mathbb{E}_x(u_x)$, see Figure~\ref{fig:profiles} a).

Both analysis lead to the same results as those obtained on the increments of the channel flow turbulent velocity, see Figure~\ref{fig:ChannelJHU_H-K}. The only difference comes from the behavior of the entropy of $\delta_{lx,ly} \left( \delta_{lx,ly} u_x\right)$ and $\tilde{u}(x,y)$ which is now independent of $\theta$ and matches the Kolmogorov theory of fully developed turbulence: a slope close to $1$ in the dissipative domain and $1/3$ in the inertial one. In particular, along the directions perpendicular to the walls, $\theta=\pm \frac{\pi}{2}$, we recover a slope $1/3$ instead of the slope $1/2$ obtained in Figure~\ref{fig:ChannelJHU_H-K} (1,c). Comparing the results obtained with $\delta_{lx,ly} \left( \delta_{lx,ly} u_x\right)$ and $\tilde{u}(x,y)$, the slope of $1$ in the dissipative domain is better recovered by the second order increments. Indeed, taking second order increments removes not only linear trends but also parabolic trends in the mean velocity profile, see Figures~\ref{fig:profiles} b and c). This de-trending is local, contrary to the removal of the mean profile.

So we conclude that for the channel flow configuration, the turbulent velocity fluctuations defined as $\tilde{u}(x,y)$ are isotropic. More generically, turbulent velocity fluctuations, once the mean velocity profile has being correctly subtracted, appear as isotropic independently of the anisotropic configuration of the flow, which only impacts the velocity profile. Increasing the order of the increments used in the multiscale decomposition allows us to appropriately remove this velocity profile and recover an isotropic behavior.

\section{Conclusions}
\label{sec:conclu}

In this paper, we presented a methodology for a multiscale nonlinear and directional characterization of images based on the estimation of high-order statistical moments and information theory quantities of the increments of the images. For a given image $I$, the entropy across scales $H_{r,\theta}(I)$ characterizes the amount of information of the increments of the process, $D_{r,\theta}(I)$ characterizes how the increment distributions are far from a Gaussian one, while the skewness $S_{r,\theta}(I)$ and flatness $F_{r,\theta}(I)$ describe respectively the asymmetry and tail prominence of the PDF of the increments of the image. We proposed a partial Theiler correction combined with random sampling methodology which, among other benefits, considerably reduces calculation time which is one of the main limitations of information theory based methods.

In order to validate our approach, we first studied 2-dimensional scale-invariant stochastic synthetic fields: fBm, a-fBm, MRW and a-MRW. We showed that our methodology correctly characterizes the monofractal behavior of both fBms and the multifractal behavior of both MRWs, allowing us to recover the scale invariant parameter $\mathcal{H}$ and the intermittency parameter $c_2$. We recovered the isotropic behavior of fBm and MRW and are able to probe the anisotropy of a-fBm and a-MRW. In particular, $F_{r,\theta}$ and $D_{r,\theta}$ are able to track the anisotropy of the intermittency parameter of the a-MRW.

A complete study of homogeneous and isotropic turbulent velocity fields allows to recover the energy distribution and cascade~\cite{Kolmogorov1991} as well as intermittency~\cite{Kolmogorov1962}. The entropy across scales characterizes the complexity of the field in the different domains of scales, while the flatness and distance from Gaussianity illustrate the multifractal nature of turbulence. Moreover, $S_{l_x,l_y}$ is different from zero along the longitudinal direction, as predicted by the $4/5$ law of Kolmogorov, while it remains close to zero in all the other directions. Thus for homogeneous and isotropic turbulence $H_{r,\theta}$, $D_{r,\theta}$ and $F_{r,\theta}$ are independent of the direction of analysis, while $S_{r,\theta}$ presents a privileged direction depending on the velocity component.

Importantly and interestingly, we tested our methodology on an inhomogeneous and anisotropic velocity field, and showed that it is able to characterize the anisotropy on several levels. On the one hand, the evolution of the entropy $H_{r,\theta}$ depends on the velocity component and the direction of analysis, and shows steeper slopes in the inertial domain for $u_x$ and $|\vec{u}|$ along the anisotropic direction of the flow, induced by the walls of the channel in the case studied here. On the other hand, intermittency characterized by $D_{r,\theta}$ and $F_{r,\theta}$ is independent of the direction of analysis and the velocity component. Finally, the estimation of $S_{r,\theta}$ is too noisy to reveal any expected anisotropy.  

Our methodology shows that turbulent velocity fluctuations can be considered isotropic, even in a strongly anisotropic and inhomogeneous setup. We show that the anisotropy of our quantities is indeed due to the inhomogeneity of the velocity field: subtracting the possibly non-linear mean velocity profile allows us to recover all results expected from Kolmogorov's fully developed turbulence theory. Additionaly, we show that using higher order increments provides another efficient way to discard locally the global trends.

Here we focused on two-dimensional fields or images, and provided a set of definitions that should prove useful in a large number of domains where images are produced daily in large quantities. However, the methodology can be straightforwardly generalized to fields of higher dimension by considering all required components in the definition of the increments.

\section*{Acknowledgment}

The authors thank the Johns Hopkins University for making available the turbulent databases.
This work was supported by the French National Research Agency (ANR-21-CE46-0011-01), within the program ”Appel \`a projets g\'en\'erique 2021”.

\bibliographystyle{elsarticle-num} 
\bibliography{THEBIBLIO}

\end{document}